\documentclass[conference,10pt]{IEEEtran}
\IEEEoverridecommandlockouts

\usepackage{algorithm}
\usepackage{graphicx}
\graphicspath{{Draft/images/}}
\usepackage{multicol}
\usepackage{lipsum}
\usepackage{mathtools}
\usepackage{amsthm}
\usepackage{amsmath,amssymb,amsfonts}
\usepackage{acronym}
\usepackage{stfloats}
\usepackage{amsfonts}
\usepackage{acronym}
\usepackage{cite}
\usepackage{multirow}
\usepackage{bm}
\usepackage{amsmath,amssymb}
\usepackage{graphicx}
\usepackage[table,xcdraw]{xcolor}
\usepackage[english]{babel}
\usepackage[caption=false,font=footnotesize]{subfig}
\usepackage[geometry]{ifsym}
\usepackage{array}
\usepackage[utf8]{inputenc}
\usepackage[T1]{fontenc}
\DeclareMathOperator*{\argmax}{arg\,max}
\usepackage{tikz}
\usepackage{lipsum}
\usetikzlibrary{calc,shadings,patterns}
\makeatletter
\tikzset{%
  remember picture with id/.style={%
    remember picture,
    overlay,
    save picture id=#1,
  },
  save picture id/.code={%
    \edef\pgf@temp{#1}%
    \immediate\write\pgfutil@auxout{%
      \noexpand\savepointas{\pgf@temp}{\pgfpictureid}}%
  },
  if picture id/.code args={#1#2#3}{%
    \@ifundefined{save@pt@#1}{%
      \pgfkeysalso{#3}%
    }{
      \pgfkeysalso{#2}%
    }
  }
}

\def\savepointas#1#2{%
  \expandafter\gdef\csname save@pt@#1\endcsname{#2}%
}

\def\tmk@labeldef#1,#2\@nil{%
  \def\tmk@label{#1}%
  \def\tmk@def{#2}%
}

\tikzdeclarecoordinatesystem{pic}{%
  \pgfutil@in@,{#1}%
  \ifpgfutil@in@%
    \tmk@labeldef#1\@nil
  \else
    \tmk@labeldef#1,(0pt,0pt)\@nil
  \fi
  \@ifundefined{save@pt@\tmk@label}{%
    \tikz@scan@one@point\pgfutil@firstofone\tmk@def
  }{%
  \pgfsys@getposition{\csname save@pt@\tmk@label\endcsname}\save@orig@pic%
  \pgfsys@getposition{\pgfpictureid}\save@this@pic%
  \pgf@process{\pgfpointorigin\save@this@pic}%
  \pgf@xa=\pgf@x
  \pgf@ya=\pgf@y
  \pgf@process{\pgfpointorigin\save@orig@pic}%
  \advance\pgf@x by -\pgf@xa
  \advance\pgf@y by -\pgf@ya
  }%
}

\makeatother

\newcounter{hatchNumber}
\usepackage{cuted}
\makeatletter
\newif\ifAC@uppercase@first%
\def\Aclp#1{\AC@uppercase@firsttrue\aclp{#1}\AC@uppercase@firstfalse}%
\def\AC@aclp#1{%
	\ifcsname fn@#1@PL\endcsname%
	\ifAC@uppercase@first%
	\expandafter\expandafter\expandafter\MakeUppercase\csname fn@#1@PL\endcsname%
	\else%
	\csname fn@#1@PL\endcsname%
	\fi%
	\else%
	\AC@acl{#1}s%
	\fi%
}%
\def\Acp#1{\AC@uppercase@firsttrue\acp{#1}\AC@uppercase@firstfalse}%
\def\AC@acp#1{%
	\ifcsname fn@#1@PL\endcsname%
	\ifAC@uppercase@first%
	\expandafter\expandafter\expandafter\MakeUppercase\csname fn@#1@PL\endcsname%
	\else%
	\csname fn@#1@PL\endcsname%
	\fi%
	\else%
	\AC@ac{#1}s%
	\fi%
}%
\def\Acfp#1{\AC@uppercase@firsttrue\acfp{#1}\AC@uppercase@firstfalse}%
\def\AC@acfp#1{%
	\ifcsname fn@#1@PL\endcsname%
	\ifAC@uppercase@first%
	\expandafter\expandafter\expandafter\MakeUppercase\csname fn@#1@PL\endcsname%
	\else%
	\csname fn@#1@PL\endcsname%
	\fi%
	\else%
	\AC@acf{#1}s%
	\fi%
}%
\def\Acsp#1{\AC@uppercase@firsttrue\acsp{#1}\AC@uppercase@firstfalse}%
\def\AC@acsp#1{%
	\ifcsname fn@#1@PL\endcsname%
	\ifAC@uppercase@first%
	\expandafter\expandafter\expandafter\MakeUppercase\csname fn@#1@PL\endcsname%
	\else%
	\csname fn@#1@PL\endcsname%
	\fi%
	\else%
	\AC@acs{#1}s%
	\fi%
}%
\edef\AC@uppercase@write{\string\ifAC@uppercase@first\string\expandafter\string\MakeUppercase\string\fi\space}%
\def\AC@acrodef#1[#2]#3{%
	\@bsphack%
	\protected@write\@auxout{}{%
		\string\newacro{#1}[#2]{\AC@uppercase@write #3}%
	}\@esphack%
}%
\def\Acl#1{\AC@uppercase@firsttrue\acl{#1}\AC@uppercase@firstfalse}
\def\Acf#1{\AC@uppercase@firsttrue\acf{#1}\AC@uppercase@firstfalse}
\def\Ac#1{\AC@uppercase@firsttrue\ac{#1}\AC@uppercase@firstfalse}
\def\Acs#1{\AC@uppercase@firsttrue\acs{#1}\AC@uppercase@firstfalse}

\renewcommand{\baselinestretch}{0.97}

\acrodef{SIC}{successive interference cancellation}
\acrodef{PAPR}{peak-to-average-power ratio}
\acrodef{APAC}{aperiodic autocorrelation}
\acrodef{OFDM}{orthogonal frequency division multiplexing}
\acrodef{DFT}{discrete Fourier transform}
\acrodef{DC}{direct current}
\acrodef{CS}{complementary sequence}
\acrodef{GCP}{Golay complementary pair}
\acrodef{ANF}{algebraic normal form}
\acrodef{PSK}{phase shift keying}
\acrodef{QAM}{quadrature amplitude modulation}
\acrodef{QPSK}{quadrature phase shift keying}
\acrodef{GDJ}{Golay-Davis-Jedwab}
\acrodef{PMEPR}{peak-to-mean envelope power ratios}
\acrodef{FFT}{fast Fourier transform}
\acrodef{BER}{bit-error ratio}
\acrodef{SNR}{signal-to-noise ratio}
\acrodef{4G}{Fourth Generation}
\acrodef{5G}{Fifth Generation}
\acrodef{NR}{New Radio}
\acrodef{LTE}{Long-Term Evolution}
\acrodef{PTS}{partial transmit sequences}
\acrodef{PSD}{power spectral density}
\acrodef{LDPC}{low-density parity check}
\acrodef{SE}{spectral efficiency}
\acrodef{eLAA}{enhanced licensed-assisted access}
\acrodef{NR-U}{NR-Unlicensed}
\acrodef{RM}{Reed-Muller}
\acrodef{AE}{autoencoder}
\acrodef{DNN}{deep neural network}
\acrodef{OFDM-AE}{OFDM-based autoencoder}
\acrodef{DL}{deep learning}
\acrodef{CP}{cyclic prefix}
\acrodef{AWGN}{additive white Gaussian noise}
\acrodef{P2C}{polar-to-Cartesian}
\acrodef{CFR}{channel frequency response}
\acrodef{ReLU}{rectified linear unit}
\acrodef{MMSE}{minimum mean sqaure error}
\acrodef{BPSK}{binary phase shift keying}
\acrodef{BLER}{block error rate}
\acrodef{ML}{machine learning}
\acrodef{PHY}{physical layer}
\acrodef{PA}{power amplifier}
\acrodef{IDFT}{inverse DFT}
\acrodef{DoF}{degrees-of-freedom}
\acrodef{IoT}{Internet-of-things}
\acrodef{M2M}{machine-to-machine}
\acrodef{DFT-s-OFDM}{discrete Fourier transform-spread orthogonal frequency division multiplexing}
\acrodef{MMSE}{minimum mean square error}
\acrodef{FDE}{frequency-domain equalization}
\acrodef{FrFT}{fractional Fourier transform}
\acrodef{TF}{time-frequency}
\acrodef{BFSK}{binary frequency-shift keying}
\acrodef{CSS}{chirp spread spectrum}
\acrodef{BCSS}{binary chirp spread spectrum}
\acrodef{EVA}{Extended Vehicular A}
\acrodef{MIMO}{multi-input multi-output}
\acrodef{PIC}{parallel interference cancellation}
\acrodef{LoRa}{Long Range}
\acrodef{HF}{high-frequency}
\acrodef{FDSS}{frequency-domain spectral shaping}
\acrodef{UAM}{urban air mobility}
\acrodef{CNS}{communication, navigation and surveillance}
\acrodef{ATM}{air traffic management}
\acrodef{ATC}{air traffic control}
\acrodef{C2}{command and control}
\acrodef{nC2}{non-command and control}
\acrodef{AI}{artifical intellegence}
\acrodef{NTN}{non-terrestrial networks}
\acrodef{UAS}{unmanned aircraft systems}
\acrodef{UTM}{UAS traffic management}
\acrodef{VFR}{visual flight rules} 
\acrodef{IFR}{instrument flight rules} 
\acrodef{RPIC}{remote pilot in command}
\acrodef{DAA}{detect \& avoid}
\acrodef{LEO}{low earth orbit}
\acrodef{MEO}{medium earth orbit}
\acrodef{GEO}{geosynchronous earth orbit}
\acrodef{SWAP}{size, weight and power}
\acrodef{CNPC}{control and non-payload communication} 
\acrodef{Con-Ops}{concept of operations}
\acrodef{RN}{relay node}
\acrodef{LOS}{line of sight}
\acrodef{BLOS}{beyond line of sight}
\acrodef{HAPS}{high-altitude platforms}
\acrodef{PTRS}{phase tracking reference symbols}
\acrodef{UE}{user equipment}
\acrodef{GSO}{geostationary synchronous orbit}
\acrodef{NGSO}{non-geostationary synchronous orbit}
\acrodef{SC}{single carrier}
\acrodef{SC-FDMA}{single carrier frequency division multiple access}
\acrodef{OBO}{out-of-band emission}
\acrodef{UL}{uplink}
\acrodef{DL}{downlink}
\acrodef{VDLM2}{VHF Data link Mode 2}
\acrodef{PA}{power amplifier}
\acrodef{PAPR}{peak-to-average ratio}
\acrodef{POT}{partially-overlapping tones}
\acrodef{FBMC}{filter-bank multicarrier}
\acrodef{FMT}{filtered multi-tone}
\acrodef{D2D}{device-to-device}
\acrodef{FO}{frequency offset}
\acrodef{WMN}{wireless mesh network}
\acrodef{UAVs}{unmanned aerial vehicles}
\acrodef{ME}{multi-user efficiency}
\acrodef{TP}{transmission point}
\acrodef{RP}{reception point}
\acrodef{RRC}{root-raised-cosine}
\acrodef{IOTA}{isotropic orthogonal transform algorithm}
\acrodef{RL}{reinforcement learning}
\acrodef{EPA}{extended pedestrian A}
\acrodef{SINR}{signal-to-interference-plus-noise ratio}
\acrodef{CCI}{co-channel interference}
\acrodef{SI}{self-interference}

\begin{document}
\title{
RL-Based Interference Mitigation in   Uncoordinated Networks with Partially Overlapping Tones
\thanks{This research is supported by the National Science Foundation (NSF) CNS through the award number 1814727.}}

\author{
  \IEEEauthorblockN{Mrugen Deshmukh\IEEEauthorrefmark{1}, Md Moin Uddin Chowdhury\IEEEauthorrefmark{1}, Sung Joon Maeng\IEEEauthorrefmark{1}, Alphan {\c{S}}ahin\IEEEauthorrefmark{2}, \.{I}smail G{\"{u}}ven{\c{c}}\IEEEauthorrefmark{1}}
    \IEEEauthorblockA{\IEEEauthorrefmark{1}Department of Electrical and Computer Engineering, North Carolina State University, Raleigh, NC, USA}
       
    \IEEEauthorblockA{\IEEEauthorrefmark{2}Electrical  Engineering Department,
University of South Carolina, Columbia, SC, USA\\
      Email: \{madeshmu,mchowdh,smaeng,iguvenc\}@ncsu.edu,~asahin@mailbox.sc.edu}
       

}

\maketitle

\begin{abstract}
\Ac{POT} are known to help mitigate co-channel interference in uncoordinated multi-carrier networks by introducing intentional \acp{FO} to the transmitted signals. 
In this paper, we explore the use of \ac{POT} with \ac{RL} in dense networks where multiple links access time-frequency resources simultaneously. 
We propose a novel framework based on Q-learning, to obtain the \ac{FO} 
for the multi-carrier waveform used for each link. In particular, we consider \ac{FMT} systems that utilize Gaussian, \ac{RRC}, and \ac{IOTA} based prototype filters. Our simulation results show that the proposed scheme enhances the capacity of the links by at least 30\% in \ac{AWGN} channel at high \ac{SNR}, and even more so in the presence of severe multi-path fading. For a wide range of interfering link densities, we demonstrate substantial improvements in the outage probability and multi-user efficiency facilitated by \ac{POT}, with the Gaussian filter outperforming the other two filters.
\end{abstract}
\acresetall

\section{Introduction}
\Ac{IoT} is a rapidly growing technology that is expected to connect billions of devices in the near future. \ac{IoT} devices can be densely located in area on the order of $10^3$ per $\text{km}^2$\cite{nokiaudn}. Hence,  they can form very dense uncoordinated networks and face spectral efficiency and throughput challenges due to complicated multi-user interference scenarios. \Ac{POT} have recently gained attention for mitigating the \ac{CCI} in uncoordinated networks. By introducing an intentional \ac{FO} for each link equal to a fraction of the frequency spacing between two subcarriers, 
\ac{POT} facilitates a reduction of multi-user interference among neighboring links.

The related research on \ac{POT} in the literature is largely based on using partially overlapping channels (using all the available spectrum) to improve throughput in wireless networks. For instance, 
partial overlapping channels are used in~\cite{ACPOCA} to improve the throughput in a remote wireless \ac{D2D} network using \ac{UAVs}, in~\cite{pot_temp2} broadly considering \acp{WMN} scenarios, and in~\cite{ICIoT} for a \ac{WMN} scenario specific to \ac{IoT} devices.
However, these studies do not consider the effect of the waveform used in the \ac{PHY}. Individual subcarriers and several schemes for waveform are considered in \cite{pot}, which lays a theoretical foundation for the analysis of using different waveform types in \ac{POT}. 
In \cite{POTGameTheory}, the authors explore \ac{POT} for the cellular networks and propose an algorithm, called \textit{Play n Wait}, to assign \acp{FO} sequentially where one user scans for the best possible \ac{FO} while all other users have to wait for several seconds. This assumption may not hold for a practical uncoordinated network. 
In this study, we focus \ac{FO} assignment 
problem for \ac{POT} as in  \cite{POTGameTheory} and aim at addressing this challenge with \ac{RL}.

In wireless networks, \Ac{RL} is commonly used to solve resource allocation \cite{wei2017user,li_drl_2019} as well as power allocation problems \cite{8422864}. For example, in \cite{wei2017user}, energy efficiency maximization problem of a hybrid-powered dense network is studied, considering an actor-critic \ac{RL} technique. In \cite{li_drl_2019}, the authors introduce \ac{RL}-based decentralized resource allocation techniques while taking strict delay constraints into account inherent in vehicle-to-vehicle networks. A distributed \ac{RL} algorithm is utilized for maintaining fairness and quality of service in dense heterogeneous networks in \cite{8422864,simsek_gc_2012,bennis_gc_2010}. In \cite{mismar_2019}, authors propose a deep Q-learning based algorithm that solves an optimization problem considering power control, beamforming, and interference coordination for sub-$6$ GHz and above-$6$ GHz bands. 

In this paper, we develop an \textit{offline} Q-learning algorithm to assign 
intentional \ac{FO} to each link. Since the algorithm is trained beforehand and available to all the users within the network, there is no time delay in assigning these \acp{FO}.
For \ac{POT}, we consider \ac{FMT}, which is a subset of filter-bank multicarrier, that allows for every subcarrier being filtered individually while allowing complex modulation symbols \cite{schaich2014waveform}, forming a grid-like structure in the time-frequency plane that partially overlapping subcarriers can exploit. 
We utilize Gaussian, \ac{RRC}, and \ac{IOTA} based prototype filters with the \ac{FMT} \ac{POT} framework, and study the associated trade-offs for using them with \ac{POT}.
With numerical results, we demonstrate the benefits of using \ac{POT} with \ac{FMT} in terms of improvements in capacity, \ac{ME}, and outage probabilities considering common propagation channels. 

The rest of the paper is organized as follows. Section \ref{sec:model} discusses the system model. Section \ref{sec:rlearning} describes our reinforcement learning based algorithm, while Section~\ref{sec:me_pot} describes the formulation of the \ac{ME} in \ac{POT}. Section~\ref{sec:results2} demonstrates the simulation results related to the proposed learning scheme. We finalize our paper in Section~\ref{sec:conclusion} with some concluding remarks.
\section{System Model}  \label{sec:model}
Consider $U$ links uniformly distributed in an area where each link consists of a \ac{TP} and a \ac{RP}. Among $U$ links, consider any one to be a link of interest with index $u$  for $u \in \{1,...,U\}$. The rest of the \acp{TP} are then considered to be \textit{aggressors} and the \ac{RP} of the link of interest is the \textit{victim}. We model the transmitted signal from the \ac{TP} of the $u$th link as
\begin{equation}
    s_u(t) = \sum_{l = -\infty}^{\infty} \sum_{n=0}^{N-1} X^u_{ln} g_{ln}(t),
    \label{eq:synthesis}
\end{equation}
where $X^u_{ln}$ is the information symbol to be transmitted from the $u$th TP, $l$ is the time index, $n$ is the subcarrier index, $N$ is the total number of subcarriers, and $g_{ln}(t)$ is the synthesis function \cite{fbmc_review} that maps $X$ to the time-frequency domain in a lattice structure as  
$g_{ln}(t) = p(t - l \tau_0)e^{j2\pi n \nu_0t}$, 
where $p(t)$ is the prototype filter being used, $\tau_0$ is the time spacing between two consecutive symbols and $\nu_0$ is the spacing between any two subcarriers. 

The received signal at the $u$th \ac{RP} can be calculated as
\begin{equation}
    y_u(t) =
    \sum_{i=1}^{U} \int_{\tau_{i,u}} h_{i,u}(\tau_{i,u},t) s_i(t-\tau_{i,u}) dt + w(t)~,
\end{equation}
where $h_{i,u}(t)$ is the channel impulse response between the desired RP of $u$th pair and TP of $i$th pair (one of which is the desired signal), $s_i(t)$ is the transmitted signal from the TP of the $i$th pair, and $w(t)$ is the \ac{AWGN}. 
The information symbol $\tilde{X}^u_{mk}$ can be obtained by calculating the projection of $y_u(t)$ onto the analysis function $\gamma_{mk}(t) =  p(t - m\tau_0)e^{j2\pi k \nu_0t}$  as \cite{fbmc_review} 
\begin{equation} \label{recX1}
    \tilde{X}^u_{mk} = \langle y_u(t),\gamma_{mk}(t) \rangle~,
\end{equation}
where  $m$ and $k$ are the time and subcarrier indices at the receiver, respectively. By grouping terms related to the interference, (\ref{recX1}) can be written as
\begin{equation} \label{recX2}
\begin{split}
    \tilde{X}^u_{mk} &= \overbrace{G_uX^u_{mk}A^u_{mkmk}}^{\text{desired symbol}} + \overbrace{G_u \sum_{l=-K+1}^{K-1} \sum_{n=0}^{N-1}X^u_{mk}A^u_{lnmk}}^{\text{self-interference}}  + \\
    & \underbrace{\sum_{i\neq u} G_i \sum_{l=-K+1}^{K-1} \sum_{n=0}^{N-1}X^i_{mk}A^i_{lnmk}}_{\text{co-channel interference}} + \underbrace{W_u}_{\text{AWGN}}~,
\end{split}
\end{equation}
where $G_u$ and $G_i$ are the channel gains at the desired signal and the $i$th aggressor, respectively, $X^i_{mk}$ is the symbol of the $i$th aggressor, $A^u_{mkmk}$ and $A^i_{nlmk}$ represent the coefficients obtained through the corresponding ambiguity functions of the desired signal and $i$th aggressor, respectively, which can be calculated as \cite{pot}:
\begin{equation}\label{eq:ambeq}
    A^i_{nlmk} = \int_{\tau} \int_{\nu}\int_t g_{ln}(t-\tau)e^{j2\pi \Delta f_i(t-\tau)}\gamma_{mk}(t)e^{j2\pi\nu t} dtd\tau d\nu,
\end{equation}
where $\Delta f_i$ is the intentional \ac{FO} given to the $i$th aggressor. 

For \ac{POT}, the amount of  \ac{CCI} and the amount of \ac{SI} in (\ref{recX2}) are adjusted though $\Delta f_i$. By sacrificing the orthogonality of the pulses on the desired link through more time-dispersive filters, an intentional \ac{FO} prevents aggressors' transmit pulses from fully overlapping with the receiver filters. 
Hence, it can reduce \ac{CCI} even if the pulses are not aligned in time, which results in higher throughput for TP-RP links. For a large network, the assignment of the \acp{FO} for each link should be chosen such that the capacity of the entire network should be optimized. Also, the parameters of the filters need to be optimized as they also affect the \ac{SI}, as can be seen from (\ref{eq:ambeq}). 
\subsection{Prototype Filters} \label{subsec:filters}
The filters used in \eqref{eq:synthesis} and \eqref{recX1} determine the inter-symbol interference characteristics. In this study, we consider three filters for analysis: 1) \Ac{RRC} filter with a roll-off factor $\alpha$, 2) Gaussian filter with a  time-frequency dispersion parameter $\rho$, and  3) \Ac{IOTA} filter with a dispersion parameter $\rho$. While \ac{RRC} filters give a set of orthogonal functions, which is a commonly used for single-carrier schemes, Gaussian pulses yield a non-orthogonal base where the basis functions are localized in both time and frequency optimally. \ac{IOTA} filter is derived from the Gaussian filter. A Gaussian filter is modified such that when it is orthogonal to its shifted versions in time and frequency. Therefore, its time-frequency characteristics are similar to Gaussian filter \cite{fbmc_review}. 

\section{Combining Q-Learning with POT} \label{sec:rlearning}

\def\numberOfAggressors[#1]{S_{#1}}
In this study, we consider Q-learning to address the \ac{FO} assignment problem. 
Q-learning is a model-free RL algorithm whose learned decision policy is determined by state-action value function $Q$ \cite{watkins1992q}, which estimates long-term discounted rewards for each state-action pair. We assume that the pairs in the network do not communicate with each other since we consider an uncoordinated network. We also assume that once an aggressor TP-RP pair enters the network, they stay active. 
Let $\numberOfAggressors[u]$ be the number of  aggressors adjacent to the $u$th link. 
The parameter $\numberOfAggressors[u]$ is estimated though a simple counting method, i.e.,  $\numberOfAggressors[u]$ increases by 1 if the \ac{SINR} of the $u$th link drops by more than 3~dB or decreases by 1 if the \ac{SINR} increases by 3~dB. 
Each pair actively monitors $\numberOfAggressors[u]$.

For the sake of illustrating the proposed scheme, consider a network with $U=3$ links. When the first link is established, there is no aggressor.
After the first aggressor enters the network, it will fully overlap with the victim at the first link. In this case, \acp{TP} at the victim link increases its counter by 1 (if the \ac{SINR} drops by more than 3 dB). The aggressor senses transmission at this desired frequency and it also increases its count by 1. At this point, the first aggressor knows that it is the second to enter the network and it takes the \ac{FO} provided by the Q-learning algorithm for $\numberOfAggressors[{u=2}]=1$. When the second aggressor enters, it can again fully overlap with the victim at $(u=1)$th link. Now, the number of aggressors from the perspective of the second aggressor is $\numberOfAggressors[{u=3}]=1$. Hence, it takes the same \ac{FO} as the first aggressor, which results in both aggressors fully overlapping. Therefore, both aggressors increase their counters, i.e., $\numberOfAggressors[{u=2}]=2$ and $\numberOfAggressors[{u=3}]=2$ and infer that there are two aggressors in the network. Since the victim has fully overlapped with both aggressors, its count will also be $\numberOfAggressors[{u=1}]=2$. The trained Q-table is stored in each \ac{TP}-\ac{RP} pair. Therefore, each link looks for the \ac{FO} values that avoid full overlapping for the new count. Through this sequence, the aggressors will also keep in memory the order in which they enter the network to ensure they do not pick the same offset provided by the proposed algorithm.  

The Q-learning algorithm is trained separately for different values of $\numberOfAggressors[u]$ in this study, adding another dimension to the Q-table. For every $\numberOfAggressors[u]$ value, all the aggressors choose from the set of possible actions and observe the corresponding rewards. The algorithm goes through a state-action value iteration process dictated by (\ref{eq:upQ}) and computes the optimal value of each state-action pair.

The states in our Q-learning algorithm are the current \ac{FO}s for all the aggressors. There are two possible actions that an agent can take in every iteration during training - \textit{$1)$ Change \ac{FO}}: The \ac{RL} algorithm can change the \ac{FO} by a fraction of the frequency carrier spacing and \textit{$2)$ Change dispersion parameter}: choose the filter dispersion parameter to reduce the interference. In this paper, we focus on changing the \ac{FO}, while the use of the dispersion parameter in the action space is left as a future study. In each time step, the agent updates the $Q(s,a)$ value, where $s$ and $a$ are one of the possible states and actions respectively, by recursively discounting future rewards and weighing them by a positive learning rate $\beta$:
\begin{equation} \label{eq:upQ}\small
    Q_{\text{new}}(s_t,a_t) \leftarrow (1-\beta)Q_{\text{old}}(s_t,a_t) + \beta [r + \gamma \max\limits_{a'\in \mathcal{A}} Q_{\text{old}}(s_{t+1},a')],
\end{equation}
where $s_t$ and $s_{t+1}$ are the current and next state, respectively, $a_t$ is the action taken at state $s_t$, $r$ is the reward for taking action $a_t$ at state $s_t$, $\gamma\in [0,1)$ is the discount parameter, and $\mathcal{A}$ is the set of possible actions. After the training process, the algorithm converges to optimal Q-values for each state-action pair~\cite{watkins1992q}, $Q^*(s, a)$ and the optimal policy can be obtained by acting greedily in every state $s$ as
\begin{equation}
\label{eq:Update_eqn}
\pi^*=\argmax\limits_{a \in \mathcal{A}} Q^*(s,a).
\end{equation}


In this study, the reward $r$ is defined as the improvement in capacity due to the change in \ac{FO} and/or the filter parameter, which can be expressed as
\begin{equation}
    r = \lambda_1(C_t - C_{t-1}),
\end{equation}
where $C_t$ is the current sum-capacity of the entire channel calculated during simulations, $C_{t-1}$ is the sum capacity in the previous iteration, and $\lambda_1$ is a weight parameter that can be tuned. The change in the sum capacity after an iteration may be very small and $\lambda_1$ allows us to amplify it for faster convergence.
The Q-value at every state is the current capacity of the desired user, given the current \ac{FO} and the filter parameters for the aggressors. After training, $u$th TP-RP pair can look up the ideal \ac{FO} and filter parameter based on $\numberOfAggressors[u]$. 

\section{Multi-user Efficiency in POT} \label{sec:me_pot}
The \ac{ME} is a measure of the signal quality over the total interference in the network. In \cite{verdu1998multiuser}, \textit{asymptotic} \ac{ME} for the desired signal $u$ is defined as
    $\eta_u \triangleq \lim_{\sigma\to\infty} \frac{e_u(\sigma)}{G_u^2}$,
where $e_u$ is the effective energy of the desired signal at the $u$th user and $\sigma$ is the standard deviation of the noise in the channel. For a conventional detector (i.e., single-user matched filter \cite{verdu1998multiuser}), $\eta_u$ is simplified to 
  $  \eta_u = \text{max}^2 \bigg\{ 0, 1 - \frac{\sum_i G_i \varrho_{i,u}}{G_u} \bigg \},$
where  $\varrho_{i,u}$ is the correlation factor between the desired user and the $i$th aggressor. For our case, the partially overlapping between the victim and aggressor links due to intentional \acp{FO} leads to correlation between them. So this intentional \acp{FO} can be considered analogous to the correlation factor. From (\ref{recX2}), the energies of the desired signal of user $u$, the \ac{SI}, and the \ac{CCI} can be calculated as
\begin{align}
    E_{\text{S}}&= G_u^2A_{mkmk}^2~,\\
    E_{\text{SI}} &= G_u^2 \sum_{l=-K+1}^{K-1} \sum_{n=0}^{N-1}A_{nlmk}^2~,\\
    E_{\text{OI}}&= \sum_iG_i^2 \sum_{l=-K+1}^{K-1} \sum_{n=0}^{N-1}A_{nlmk}^2~.
\end{align}
The effective symbol energy for the desired user after taking the interference into account is then calculated as
\begin{equation}
\small
\begin{split}
    e_u &= G_u^2A_{mkmk}^2 - G_u^2 \sum_{l=-K+1}^{K-1} \sum_{n=0}^{N-1}A_{nlmk}^2 \\
    & - \sum_iG_i^2 \sum_{l=-K+1}^{K-1} \sum_{n=0}^{N-1}A_{nlmk}^2~.
\end{split}
\end{equation}
As a result, the \ac{ME} is obtained as
\begin{equation}\label{eq:ame_final}
\eta_u  = \frac{e_u}{G_u^2} 
   = \text{max}^2 \bigg\{ 0, 1 - \frac{\sqrt{E_{\text{SI}} + E_{\text{OI}}}}{G_u A_{mkmk}}\bigg\}~.
\end{equation}

\section{Simulation Results} \label{sec:results2}
In this section, we evaluate the performance of the proposed learning method with computer simulations.
We uniformly distribute the TPs and RPs in an area of $1~\text{km}^2$ and pair them.
Considering \ac{IoT} applications, we assume a modestly small resource allocation for data with $N = K = 12$ for all links in the network. 
We use \ac{QPSK} modulation 
throughout the simulations. To calculate path loss, we use the free-space path loss model without loss of generality of the proposed scheme. The carrier frequency is $800$~MHz with the bandwidth of $200$ kHz for each TP-RP link. In current simulations, we keep the filter parameter to be constant ($\rho= \alpha= 0.2$). 
The multi-path channel is modeled based on \ac{EPA} specified in \ac{LTE} standards, unless otherwise stated.
 \begin{figure}[t]
     \centering
        \subfloat[Gaussian filter ($\rho = 1$).]{ \label{fig:ambGaussian2}
			\includegraphics[width=1.7in]{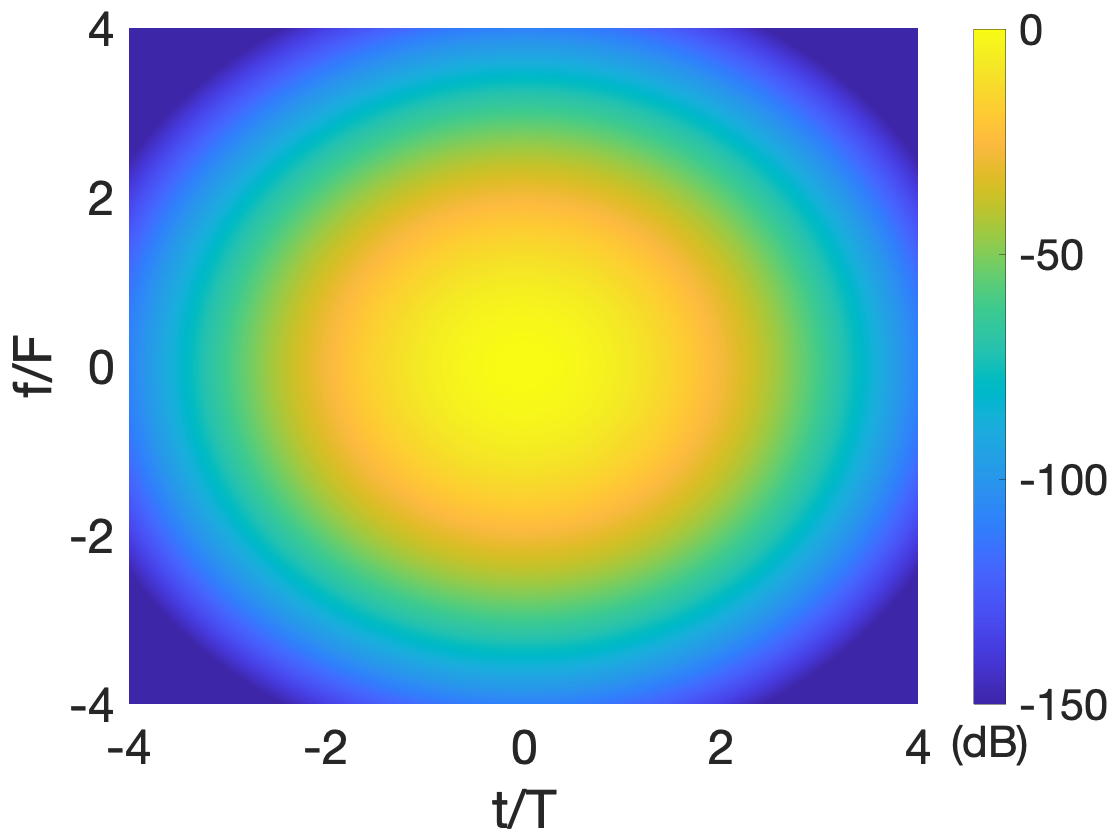}}	      
        \subfloat[RRC filter ($\alpha = 1$).]{ \label{fig:ambRRC2}
			\includegraphics[width=1.7in]{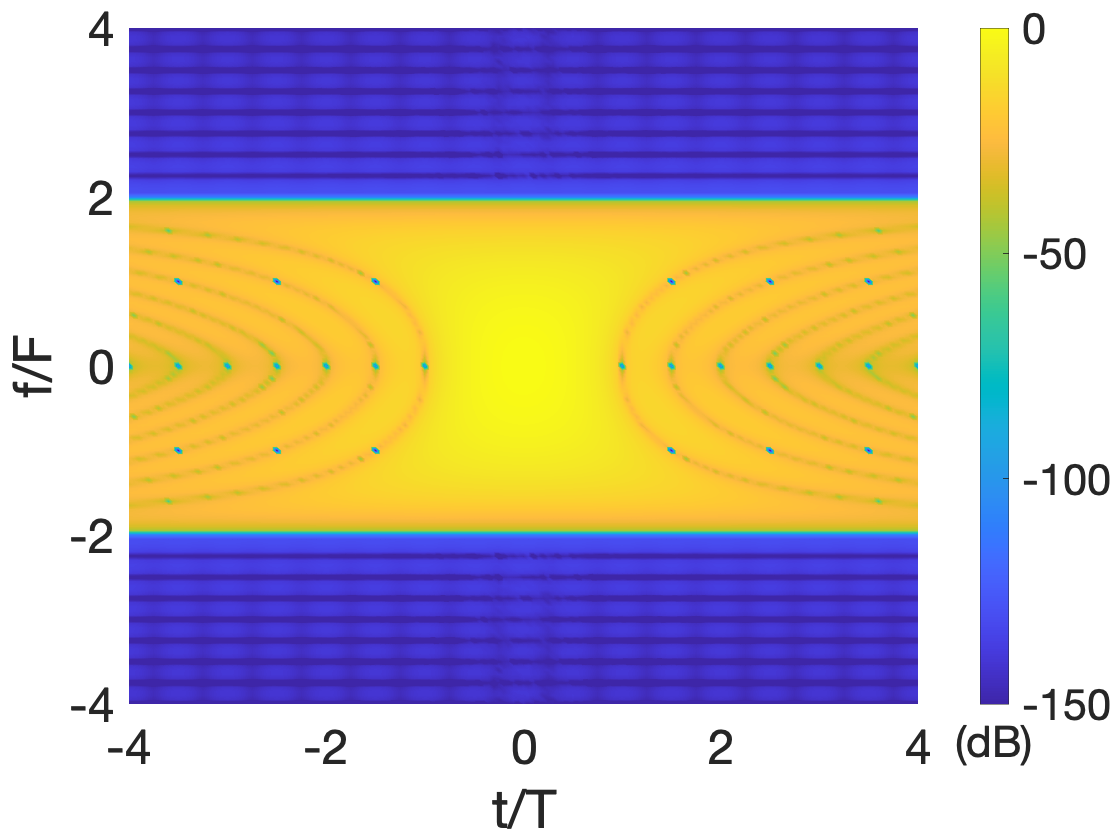}}		
\\		
        \subfloat[Gaussian filter ($\rho = 0.5$).]{ \label{fig:ambGaussian1}
			\includegraphics[width=1.7in]{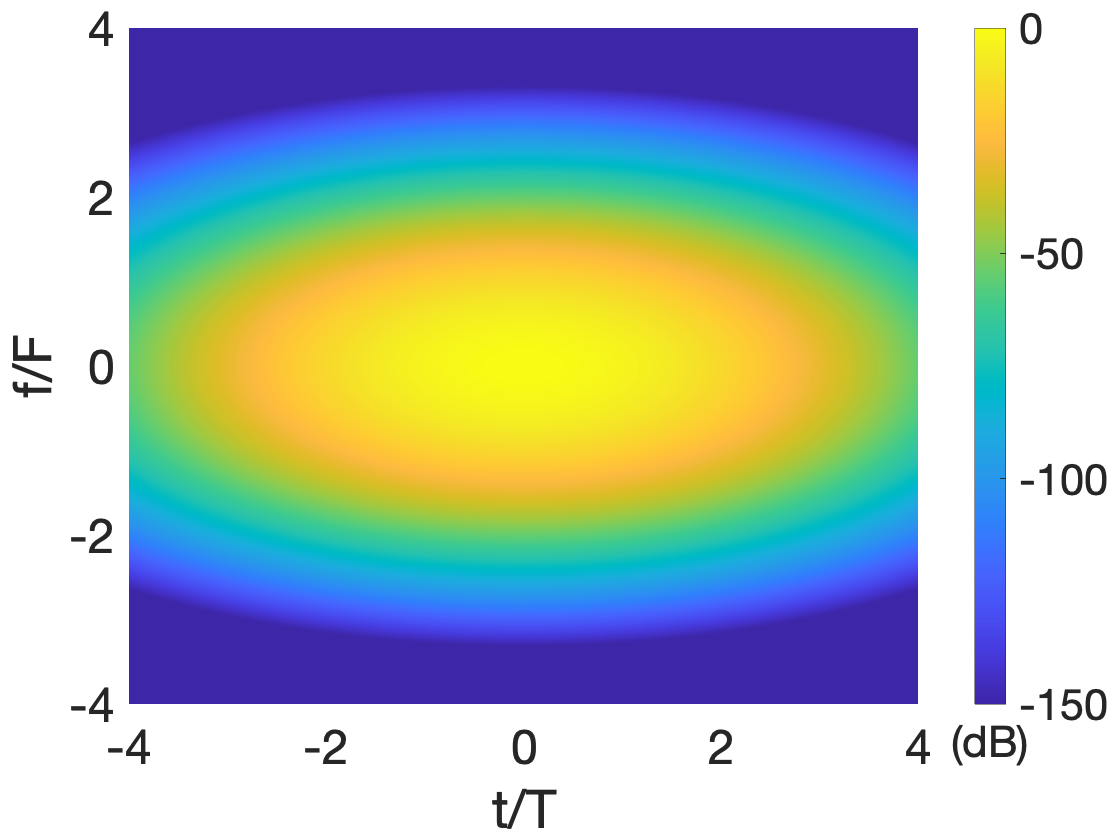}}
\subfloat[RRC filter ($\alpha = 0.5$).]{ \label{fig:ambRRC1}
			\includegraphics[width=1.7in]{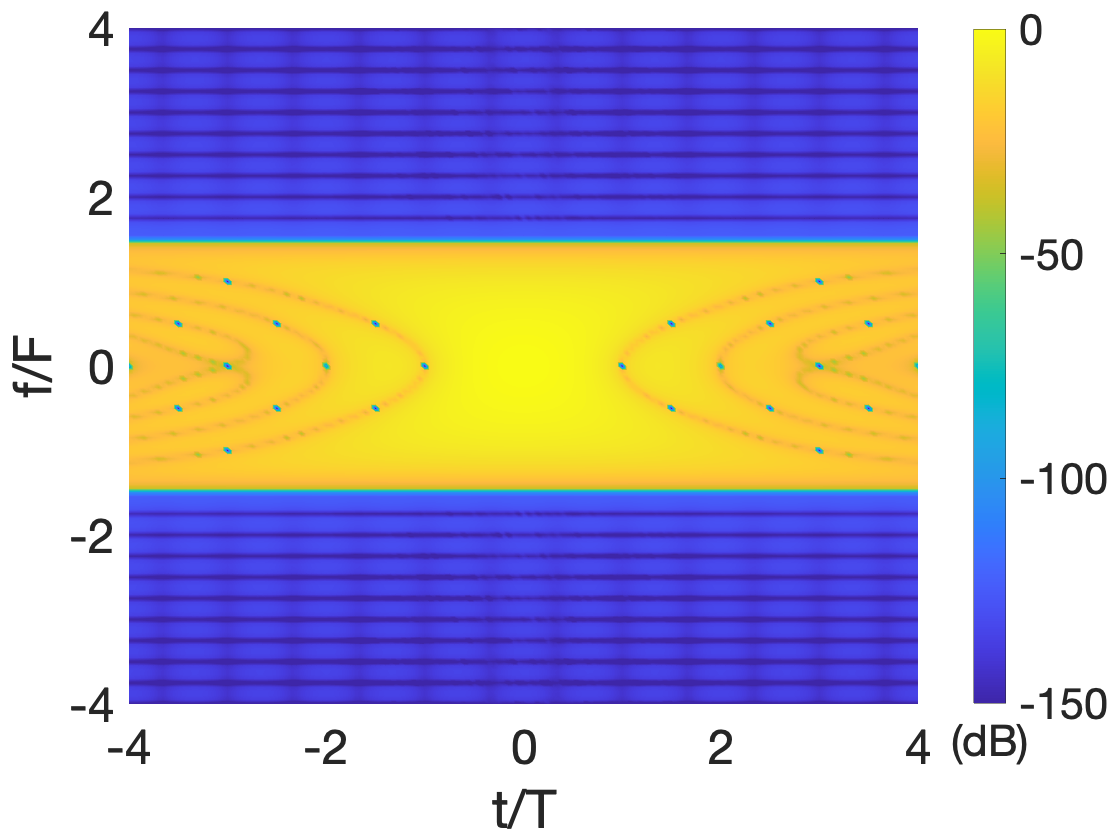}}
			\caption{The ambiguity functions of RRC and Gaussian filters for different dispersion parameters.}
			\label{fig:ambiguities}
\vspace{-3mm}
\end{figure} 

\begin{figure}[ht]
     \centering
      	\subfloat[Capacity under \ac{AWGN} channel.]{ \label{fig:cap_perfect}
			\includegraphics[width=8cm]{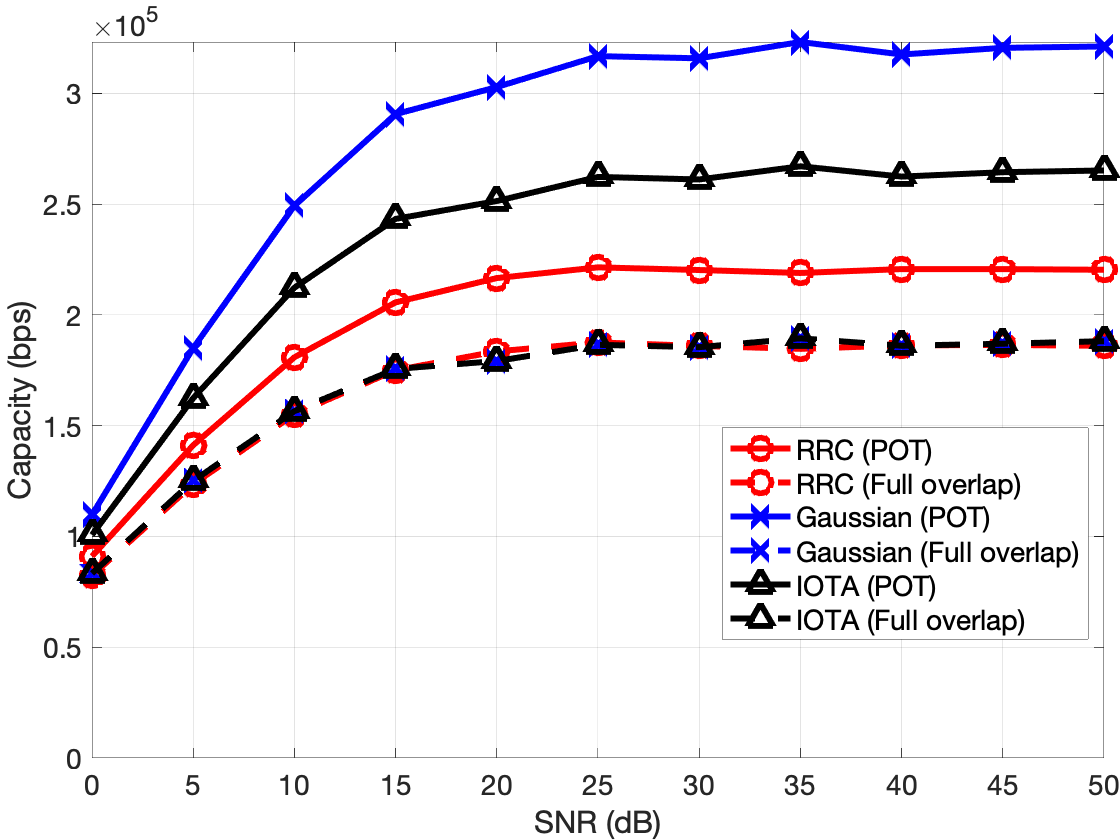}}\\
        \subfloat[Capacity under EPA channel model.]{ \label{fig:cap_epa}
			\includegraphics[width=8cm]{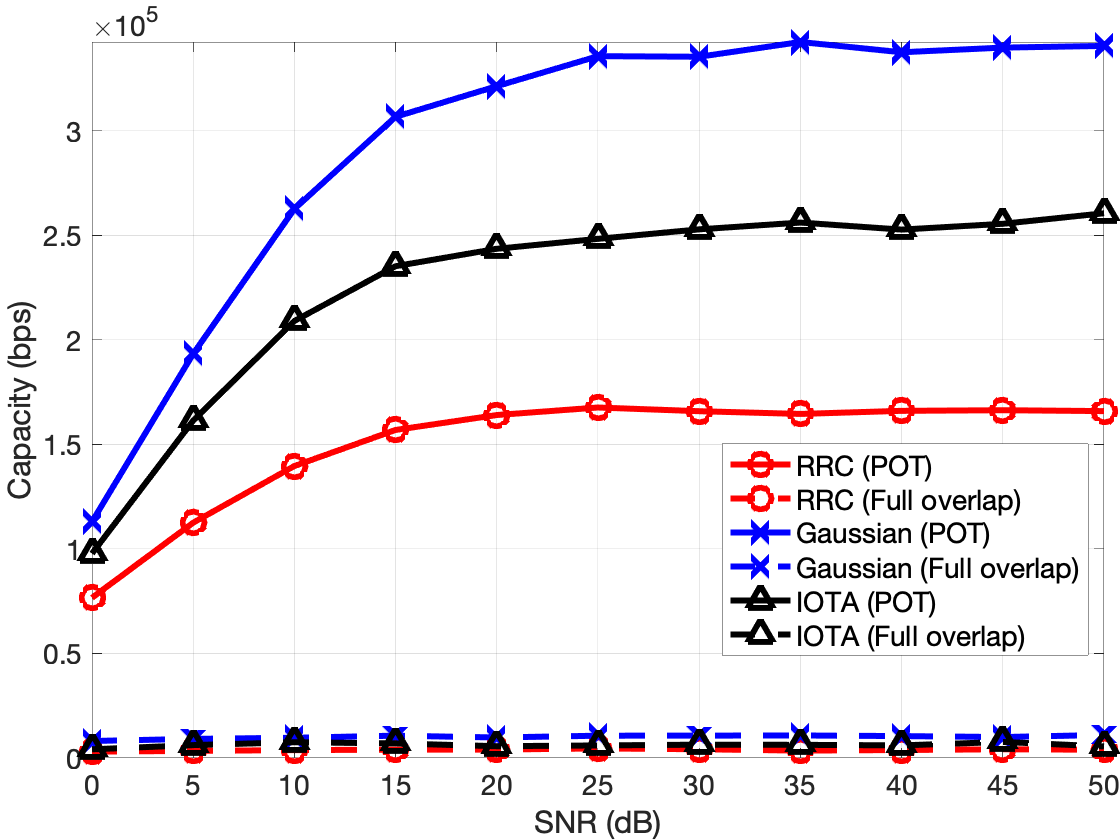}}
			\caption{Capacity analysis for different channels.}
			\label{fig:cap_vs_noise}
\vspace{-0.4cm}
\end{figure}

\subsection{Effect of Filter Parameters on Interference}
In this section, we analyze the impact of the dispersion parameter on the systems performance. In \figurename~\ref{fig:ambiguities}, we compare the ambiguity plots for \ac{RRC} and Gaussian filters for different dispersion parameters. For  Gaussian  filter, when the dispersion parameter $\rho = 1$, the energy is distributed equally in the time and frequency domain (i.e., isotropic) as in \figurename~\ref{fig:ambiguities}\subref{fig:ambGaussian2}. However, when we set $\rho = 0.5$, the filter starts to {\em squeeze} in the frequency domain and expand in the time domain as in \figurename~\ref{fig:ambiguities}\subref{fig:ambGaussian1}. Thus, in the presence of \ac{CCI}, a lower $\rho$ allows other users to exploit available responses in frequency. However, this leads to the problem of \ac{SI} at the desired link as it causes non-orthogonal pulses. We observe the same behaviour for \ac{RRC} filter in \figurename~\ref{fig:ambiguities}\subref{fig:ambRRC2} and \figurename~\ref{fig:ambiguities}\subref{fig:ambRRC1} for different $\alpha$ parameters. Dealing with this \ac{SI} requires an equalizer being used at the receiver, and is out of the scope of this paper. The behavior of \ac{IOTA} filter is similar to the Gaussian filter with more energy spreading in time and frequency. For the ambiguity function for \ac{IOTA} filter, we refer the reader to \cite{fbmc_review}.


\subsection{Capacity Analysis}
In \figurename~\ref{fig:cap_vs_noise}, we analyze the capacity at the desired link (i.e., average capacity over different instances for the same link) for different filters against when ten aggressors are present. \figurename~\ref{fig:cap_vs_noise}\subref{fig:cap_perfect} and \figurename~\ref{fig:cap_vs_noise}\subref{fig:cap_epa} show the capacity curves for \ac{AWGN} and \ac{EPA} channel models, respectively. Among the filters, Gaussian filter provide the best capacity performance in both channel condition. As compared to fully overlapping, \ac{POT} provides a gain by approximately $80\%$ with Gaussian filters without noise being present. Another important observation is that the capacity reduces without \ac{POT} under \ac{EPA} channel model, while the drop is not substantial for \ac{POT}.

\begin{figure}[t]%
    \centering
    \includegraphics[width=8cm]{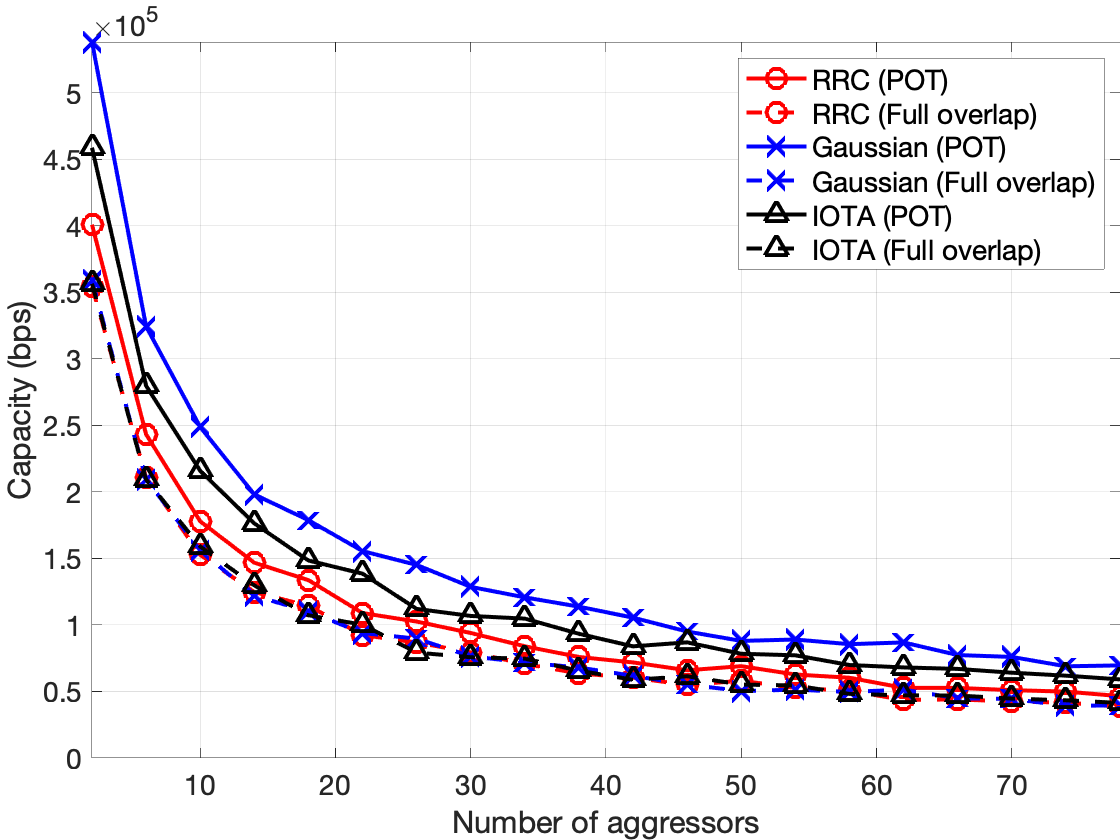}
    \caption{Capacity versus number of aggressors with POT and full overlap.}
    \label{fig:capN}%
\vspace{-3mm}
\end{figure}

In \figurename~\ref{fig:capN}, we investigate the capacity in the channel for a victim link against the number of aggressors present in the network at  $10$ dB \ac{SNR}. As expected, with the increasing interference from the number of aggressors, the capacity reduces gradually. Gaussian filter offers a small improvement over other filters when \ac{POT} are used. The capacity when full overlapping occurs is the same for any type of filter used. 
In the case of severe \ac{CCI} (e.g. $1000$ $\text{femtocells/km}^2$ \cite{PAreview}), the curves for full overlapping and partial overlapping converge.
\subsection{Multi-user Efficiency and Outage Probability}
\begin{figure}[t]%
    \centering
    \includegraphics[width=8cm]{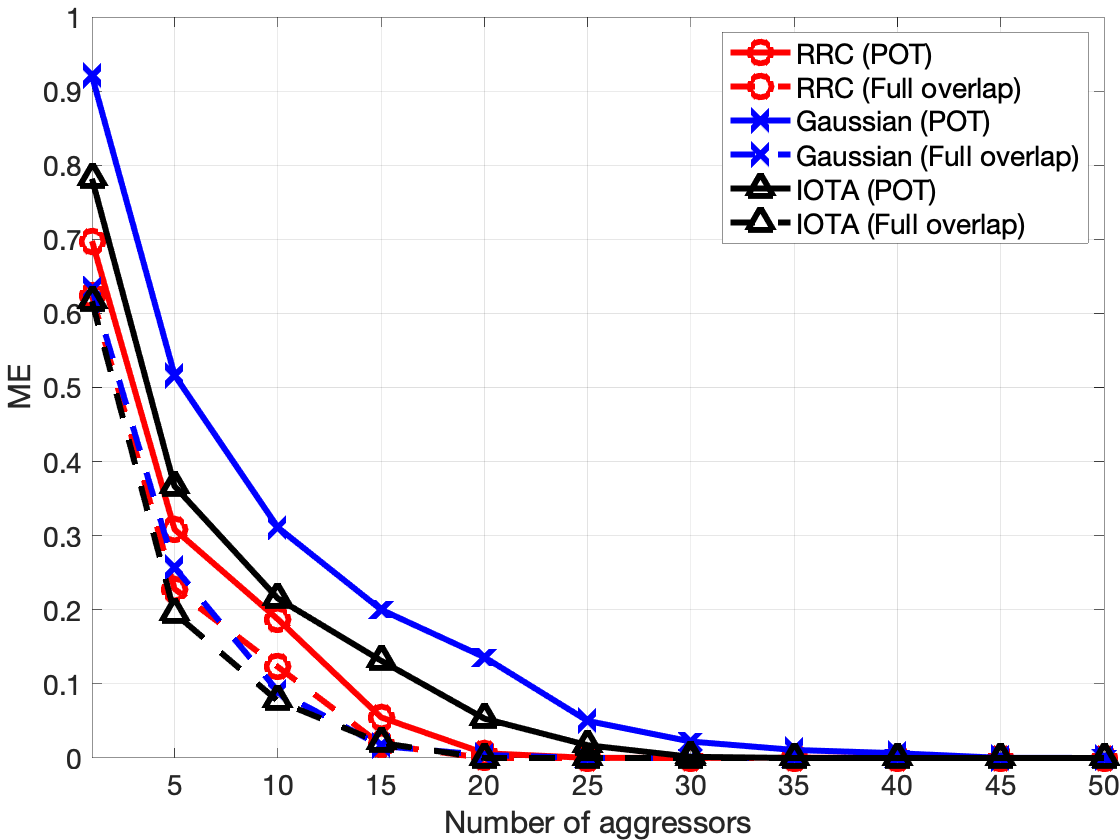}
    \caption{Multi-user efficiency with POT and full overlap.}
    \label{fig:ame_final}%
    \vspace{-3mm}
\end{figure}

\begin{figure}[t]%
    \centering
    \includegraphics[width=8cm]{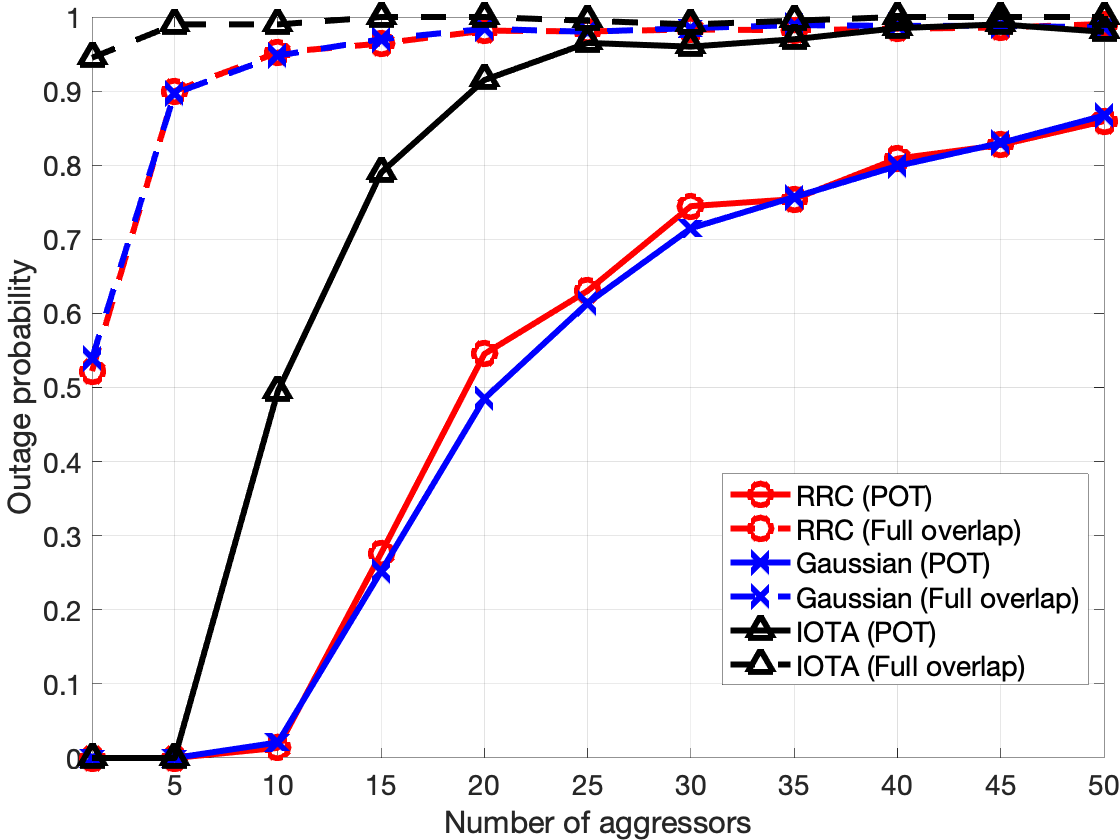}
    \caption{Outage probability  with POT and full overlap ($\Gamma = -6$ dB).}
    \label{fig:out1}%
    \vspace{-3mm}
\end{figure}

Based on \eqref{eq:ame_final}, the \ac{ME} for different filters at different interference conditions are plotted in \figurename~\ref{fig:ame_final}. Among all the filters, the Gaussian filter gives a higher \ac{ME} for \ac{POT}. For a full overlapping scenario, all filters provide similar ME.
We present the outage probabilities in \figurename~\ref{fig:out1}. The outage occurs when the \ac{SINR} of the desired signal falls below a pre-defined threshold denoted by $\Gamma$. \figurename~\ref{fig:out1} shows the outage probabilities for $\Gamma = - 6$  (specified in LTE). The Gaussian filter outperforms other filters again and shows the least probability of outage against any number of aggressors. 

\section{Conclusion and Future Work} \label{sec:conclusion}
Our preliminary results show the benefits of using \ac{POT} in uncoordinated networks. We show that users to the order of $10^2$ can be accommodated at a cost (to performance) significantly lower than the one with fully overlapping. We observe that a non-orthogonal filter (Gaussian) has the highest potential to fully exploit benefits offered by POT. Gaussian filters yield superior capacity against noise and interference, while also providing higher \ac{ME}. For outage probability, using RRC filters provides similar performance but Gaussian slightly edges out ahead.     
The primary disadvantage of the proposed scheme is that the algorithm becomes more computationally expensive with the increasing number of aggressors, requiring a longer training period. As an extension of the current study,  a detailed analysis of the computational complexity versus the accuracy of training is needed. Another extension is an \textit{online} Q-learning algorithm which accounts for cases where a TP-RP pair has a wrong estimate of the number of aggressors. Incorporating an equalizer which addresses the inherent \ac{SI} for non-orthogonal filters also needs to be investigated to improve the link performance further.   

\bibliographystyle{IEEEtran}
\bibliography{ref}

\end{document}